\documentclass{ijcsa}
\usepackage{amssymb}
\usepackage{graphicx}
\usepackage{epstopdf}
\begin{document}
\catchline{}{}{}{}{}

\markboth{Dey, Al-Qaheri, Sane, Sanyal}
{Bounds for the Generalized Fibonacci-p-Sequence and its Application in Data-Hiding}

\title{A Note On the Bounds for the Generalized Fibonacci-p-Sequence and its Application in Data-Hiding}

\author{Sandipan Dey
\\Microsoft India Development Center
\\\emph{sandipan.dey@gmail.com}
}
\author{Hameed Al-Qaheri
\\Department of Quantitative Methods and Information Systems
\\College of Business Administration
\\Kuwait University
\\\emph{alqaheri@cba.edu.kw}
}
\author{Suneeta Sane
\\Computer and Information Technology Department
\\Veermata Jijabai Technological Institute
\\Mumbai, Maharashtra 400019, India
\\\emph{sssane@vjti.org.in}
}
\author{Sugata Sanyal
\\School of Technology and Computer Science
\\Tata Institute of Fundamental Research
\\Homi Bhabha Road, Mumbai - 400005, India
\\\emph{sanyal@tifr.res.in}
\\July 2008
}

\date{}

\maketitle

\begin{abstract}
\par In this paper, we suggest a lower and an upper bound for the Generalized Fibonacci-p-Sequence, 
for different values of $p$. The Fibonacci-p-Sequence is a generalization of
the Classical Fibonacci Sequence. We first show that the ratio of two consecutive terms in generalized 
Fibonacci sequence converges to a p-degree polynomial and then use this result to prove the  
bounds for generalized Fibonacci-p sequence, thereby generalizing the exponential bounds for 
classical Fibonacci Sequence. Then we show how these results can be used to prove efficiency for 
data hiding techniques using generalized Fibonacci sequence. These steganographic techniques 
use generalized Fibonacci-p-Sequence for increasing the number of available bit-planes to hide data, 
so that more and more data can be hidden into the higher bit-planes of any pixel without causing much 
distortion of the cover image. This bound can be used as a theoretical 
proof for efficiency of those techniques, for instance it explains why more and more data can be 
hidden into the higher bit-planes of a pixel, without causing considerable decrease in PSNR. 

\end{abstract}

\section{Introduction}
\par 

Among many different data hiding techniques proposed to embed secret message
within images, the LSB data hiding technique is one of the simplest methods
for inserting data into digital signals in noise free environments, which merely
embeds secret message-bits in a subset of the LSB planes of the image. 
(LSB is the least significant bit or the 0th bit, the second LSB is the 1st bit, and so on). 
Despite being simple, this technique is more predictable and hence less secure, 
also PSNR (peak signal to noise ratio) decreases very rapidly as we use the higher bit planes for data hiding. 
As soon as we go from LSB (least significant bit) to MSB (most significant bit) for selection of bit-planes for
our message embedding, the distortion in stego-image is likely to increase
exponentially, so it becomes impossible (without noticeable distortion and
with exponentially increasing distance from cover-image and stego-image)
to use higher bit-planes for embedding without any further processing.
The workarounds may be: through the random LSB replacement (in stead
of sequential), secret messages can be randomly scattered in stego-images, so
the security can be improved. Also, using the approaches given by variable depth LSB algorithm \cite{liu04}, 
or by the optimal substitution process based on genetic algorithm and
local pixel adjustment \cite{wang01}, one is able to hide data to some extent
in higher bit-planes as well.\\

Battisti {\it et al.} \cite{battisti06}, \cite{pic06} proposed a novel data hiding technique from a totally different perspective, it uses a different bit-planes decomposition altogether, based on the generalized Fibonacci-p-sequences, that not only increases the number of embeddable bit-planes but also decreases PSNR in the stego image considerably, thereby improving the LSB technique. In this paper, we first prove some theoretical upper and lower bounds for generalized Fibonacci-p-sequence and give a theoretical proof for the better performance of the data hiding technique using generalized Fibonacci decomposition, i.e., why the data hiding technique using this decomposition not only gives larger number of bit planes for hiding secret bits, but also gives a far better PSNR than that in classical LSB technique. \\

\par The Generalized Fibonacci-p-Sequence (\cite{horadam61}, \cite{basin63}, \cite{hoggatt72}, \cite{atkins87}, \cite{hendel94}, \cite{sun92}) is given by, 

\begin{eqnarray}
\label{fib}
\nonumber F_{p}(0)=F_{p}(1)=\ldots=F_{p}(p)=1,\\
F_{p}(n)=F_{p}(n-1)+F_{p}(n-p-1),\;\forall{n\geq p+1},\;\;n,p\in\mathbb{N}
\end{eqnarray}

\par For $p=1$, we have,
\begin{eqnarray*}
F(0)=F(1)=1,\\
F(n)=F(n-1)+F(n-1),\;\forall{n\geq2},\;n \in \mathbb{N}
\end{eqnarray*}
\par We get the classical Fibonacci sequence $1,1,2,3,5,8,\ldots$. We already have some results for this
Classical Fibonacci Sequence, e.g., we know the ratio of two consecutive terms in Fibonacci sequence converge to Golden Ratio,  
$\frac{1+\sqrt{5}}{2}$. 
In this paper, we show that $\alpha_{p}^{n}>F_{p}(n)>\alpha_{p}^{n-p},\;\forall{n,p} \in \mathbb{N}$, where $\alpha_{p}$ is the positive Root of $x^{p+1}-x^{p}-1=0$. The ratio of two consecutive terms in Fibonacci-p-Sequence converges to this $\alpha_{p}$.

\section{Bounds for the generalized Fibonacci-p-Sequence}
In this section we prove the upper and lower bounds for the generalized Fibonacci-p-sequence. First, we show that the ratio of 
consecutive terms of generalized Fibonacci-p-sequence converges to the positive root of the polynomial $x^{p+1}-x^p-1=0$. Next 
we use this root to prove a bound on the generalized Fibonacci-p-sequence.

\subsection{Lemma 1}
\label{sec:Lemma1}
\par The ratio of two consecutive numbers in generalized Fibonacci p-sequence converges to the positive root of 
the degree-$p$ polynomial $P(x)=x^{p+1}-x^{p}-1$.

\paragraph{Proof:}
\label{sec:Proof}

\subsubsection*{Convergence}
Let us first define the ratio of two consecutive terms of Fibonacci-p-sequence as a sequence $\{\beta_{n}\}=\left\{\frac{F_{p}(n+1)}{F_p(n)}\right\}$. \\
Now, by definition of Fibonacci-p-sequence, we have
\begin{eqnarray*}
{\beta}_0={\beta}_1=\cdots={\beta}_p=1 \\ 
{\beta}_n=1+\frac{F_{p}(n-p)}{F_p(n)}>1, \; \forall{n}>p \\
{\beta}_{p+1}={\beta}_p+{\beta}_0=1+1=2\\
{\beta}_n=1+\frac{F_{p}(n-p)}{F_p(n)} < 1 + 1 = 2, \; \forall{n}>p+1 \\
\Rightarrow 1 < {\beta}_n < 2 , \; \forall{n}>p+1
\end{eqnarray*}

We observe that the sequence ${\beta}_n$ is bounded and hence by Monotone Convergence theorem must 
have a convergent subsequence. For instance, for $p=1$ (classical Fibonacci sequence), we have two convergent subsequences
${\beta}_{2n}$ (increasing) and ${\beta}_{2n+1}$ (decreasing), $n \in \mathbb{N}$ (natural number) and they both converge to the same limit
$\frac{1+\sqrt{5}}{2}\approx 1.618$ \cite{craw00}, as shown in figure 2.  

\subsubsection*{Positive Root of the polynomial $x^{p+1}-x^{p}-1$}
Now, let's analyze the polynomial function $y = P(x) = x^{p+1}-x^{p}-1$. By Descartes' rule, the polynomial
can have at most one positive real root, since it has exactly one change in sign. \\

First we observe that the function $P(x)$ is continuous and differentiable everywhere. We also notice that the function has exactly one positive root $\alpha_{p}$ (and $\alpha_{p}$ is also strictly larger than 1). This is a consequence of elementary calculus. 
By successive differentiation, we see that 

\begin{eqnarray*}
y_{1} = P'(x) = (p+1)x^{p}-px^{p-1} \\	
y_{2} = P''(x) = (p+1)px^{p-1}-p(p-1)x^{p-2} \\
\end{eqnarray*}

\par 
\begin{itemize}
	\item The function $P(x)$ has critical points at $P'(x)=0$, i.e., at $x=0$ and $x=\frac{p}{p+1}$.
	\item ${y_{2}\Big|}_{x=\frac{p}{p+1}}=\frac{p^{p-1}}{(p+1)^{p-2}}>0,\;\forall{p}\geq{1}$.
	\item By 2nd order sufficient condition for local minima, $P(x)$ has a (local) minima at $x=\frac{p}{p+1}$.
	\item At $x=0$, the function will have a maxima or a point of inflection depending on whether $p$ is odd or even,
				explained in the next section.
\end{itemize}

When $p \in \mathbb{N}_{odd}$, we have,
		\par $y_{1} = P'(x) =(p+1)x^{p-1}\left(x-\frac{p}{p+1}\right)\; \mbox{is} \left\{
		\begin{array}{c l}
			<0 & x < 0 \\
			=0 & x = 0 \\
			<0 & 0 < x < \frac{p}{p+1} \\
			=0 & x = \frac{p}{p+1} \\
			>0 & x > \frac{p}{p+1}
		\end{array}
		\right.$ \\
		
Hence the function is decreasing in $\left(-\infty, \frac{p}{p+1}\right)$ and increasing in $\left(\frac{p}{p+1}, \infty
\right)$.	Also, $y(-1)=1$,  $y(0)=y(1)=-1$  and $y(2)=2^{p}-1\geq 1, \; \forall{p} \in \mathbb{N}_{odd}$. At $x = 				
\frac{p}{p+1}$, $P(x)$ has a minima (gradient changes from $negative$ to $positive$) but at $x=0$ (no sign change in gradient) we 
have a point of inflection. Again, $P(x)$ being a continuous function assumes all possible values within an interval. 
Combining all these, we can easily see that the graph of the function has exactly two real zeroes, 
one positive and the other negative (the remaining roots are complex conjugate pairs). \\

When $p \in \mathbb{N}_{even}$, we have,
	\par $y_{1} = P'(x) =(p+1)x^{p-1}\left(x-\frac{p}{p+1}\right)\; \mbox{is} \left\{
		\begin{array}{c l}
			>0 & x < 0 \\
			=0 & x = 0 \\
			<0 & 0 < x < \frac{p}{p+1} \\
			=0 & x = \frac{p}{p+1} \\
			>0 & x > \frac{p}{p+1}
		\end{array}
		\right\}$ \\
		
		Hence the function is increasing in $(-\infty, 0)$, then decreasing in $\left(0, \frac{p}{p+1}\right)$ and again 
		increasing in $\left(\frac{p}{p+1}, \infty \right)$.	Also, $y(-1)=y(0)=y(1)=-1$  and $y(2)=2^{p}-1\geq 1, \; \forall{p} 	
		\in \mathbb{N}_{even}$. At $x_{0} = \frac{p}{p+1}$, $P(x)$ has a minima but at $x=0$ (gradient changes from $positive$ to $negative$) we 
		have a maxima. Combining all these, we can easily see that we have exactly one real (positive) root at $\alpha_{p}$, since
		$y(1)<0$ and $y(2)>0$, also $y(x)$ being continuous. (From this result we immediately have Lemma 2). From figure 1		
		we can see the graph of the degree-$p$ polynomial, for odd and even $p$ respectively.

\subsubsection*{Convergence to the positive root of the polynomial}
Now, we show that if sequence \{$\beta_{n}$\} converges to $\beta$, then $\beta=\alpha_{p}$. \\

Let's assume the sequence \{$\beta_{n}$\} converges to $\beta \in \Re^{+}$. Now we prove, the sequence must converge to the only positive root $\alpha_{p}$ of the above-stated $p$-degree polynomial. By assumption,
\begin{eqnarray*}
\beta=\lim_{n \rightarrow \infty}{\left(\frac{f_{n+p}}{f_{n+p-1}}\right)}= \lim_{n \rightarrow \infty}{\left(\frac{f_{n+p-1}}{f_{n}}\right)}
= \ldots = \lim_{n \rightarrow \infty}{\left(\frac{f_{n}}{f_{n-1}}\right)}=\ldots,\\
f_{n}=n^{th}\;number\;in\;the\;Fibonacci-p\;Sequence,\; f_{n+p}=f_{n+p-1}+f_{n-1}\\
\Rightarrow \beta=\lim_{n \rightarrow \infty}{\left(\frac{f_{n+p-1}+f_{n-1}}{f_{n+p-1}}\right)}
= \lim_{n \rightarrow \infty}{\left(\frac{f_{n}}{f_{n-1}}\right)},\;\\
\Rightarrow \beta
=1+\lim_{n \rightarrow \infty}{\prod_{k=n-1}^{k=n+p-2} {\left(\frac{f_{k}}{f_{k+1}}\right)}}= \lim_{n \rightarrow \infty}{\left(\frac{f_{n}}{f_{n-1}}\right)}	\\
\Rightarrow \beta=1+\prod_{k=1}^{k=p}\left(\frac{1}{\beta}\right)\;\Rightarrow \beta=1+\frac{1}{\beta^{p}}\\
\Rightarrow \beta^{p+1}-\beta^{p}-1=0 \\
\label{lemma1Proof}
\end{eqnarray*}

Hence $\beta$ satisfies the equation $x^{p+1}-x^{p}-1=0$, $\forall{p} \in \mathbb{N}$, $\beta \in \Re^{+}$, i.e.,
from above results, we have, $\beta = \alpha_{p}$

\begin{figure}[htbp]
	\label{fig:f1}
	\centering
		\includegraphics[width=12cm,height=6cm]{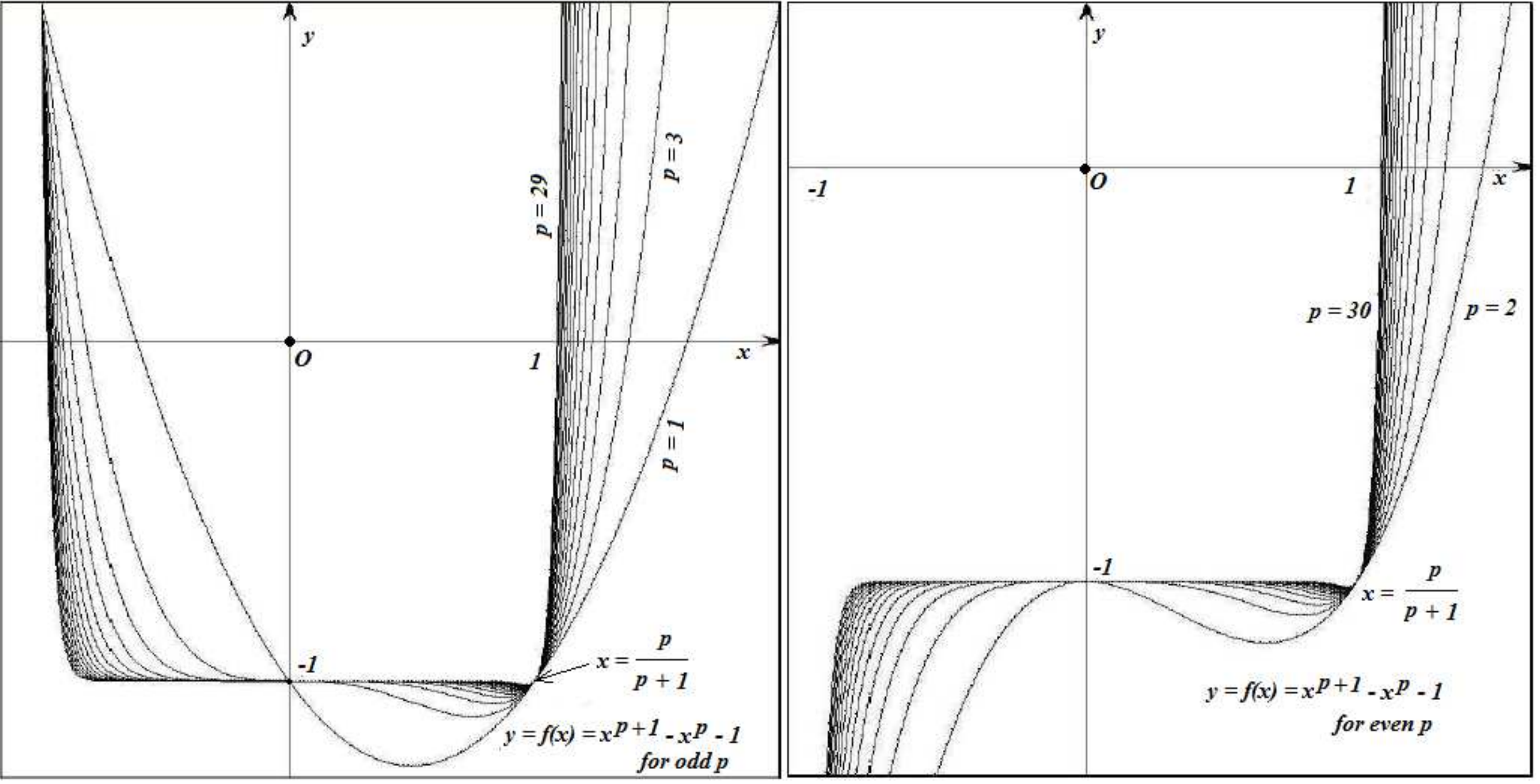}
	\caption{Graph of $x^{p+1}-x^{p}-1=0$, showing $\alpha_p$ for different values of $p$}
\end{figure}

\subsection{Lemma 2}
\label{sec:Lemma2}
\par If $\alpha_{p}$ be a positive root of the equation $x^{p+1}-x^{p}-1=0$, we have $1<\alpha_{p}<2$,
$\forall{p} \in \mathbb{N}$.
\paragraph{Proof:}
\label{sec:Proof_2}
We have,
\begin{eqnarray}
\label{ineq_p}
\alpha_{p}^{p+1}-\alpha_{p}^{p}-1=0	\nonumber \\
\mbox{Also,} \;\;\; 2^{p+1}-2^{p}-1=2^{p}-1>0, \;\forall{p} \in \mathbb{N} \nonumber \\
\Rightarrow 2^{p}-1>\alpha_{p}^{p+1}-\alpha_{p}^{p}-1 \nonumber \\	
\Rightarrow (2^{p}-\alpha_{p}^{p})>\alpha_{p}^{p}(\alpha_{p}-2)
\end{eqnarray}
Also,
\begin{eqnarray}
\label{ineq_p_1}
-1<0=\alpha_{p}^{p+1}-\alpha_{p}^{p}-1 \nonumber \\	
\Rightarrow \alpha_{p}^{p}(\alpha_{p}-1)>0 \nonumber \\	
\Rightarrow \alpha_{p}>1 \;\;\mbox{(since positive)}
\end{eqnarray}

\par From (\ref{ineq_p}), we immediately see the following:
\begin{itemize}
	\item $\alpha_{p}>0$ according to our assumption, hence we can not have $\alpha_{p}=2$ 
				(LHS \& RHS both becomes 0, that does not satisfy inequality (\ref{ineq_p})).
	\item If $\alpha_{p}>2$, we have LHS $< 0$ while RHS $>0$ which again does not satisfy inequality (\ref{ineq_p}).
	\item Hence we have $\alpha_{p}<2,\;\forall{p}\in\mathbb{N}$
\end{itemize}
\par From (\ref{ineq_p_1}), we have, $\alpha_{p}>1$.
Combining, we get, $1<\alpha_{p}<2,\;\forall{p}\in\mathbb{N}$

\subsection{Lemma 3}
\label{sec:Lemma3}
\par If $\alpha_{p}$ be a positive root of the equation $x^{p+1}-x^{p}-1=0$, where $p \in \mathbb{N}$, we have the following results, $\forall{k}\in\mathbb{N}$,
\begin{itemize}
\item	$\alpha_{k}>\alpha_{k+1}$
\item	$\lim_{k \rightarrow \infty}{\alpha_{k}}=1$ 
\item	$\alpha_{k+1}>\frac{1+\alpha_{k}}{2}$
\item	$\alpha_{k}^{k}<(k+1)$
\item	$\alpha_{k}^{k+1}>1$
\end{itemize}
\paragraph{Proof:}
\label{sec:Proof_3}
We have,
\begin{eqnarray}
\label{eq_p}
\mbox{For } p=k, \; \alpha_{k}^{k+1}-\alpha_{k}^{k}-1=0	\nonumber \\
\mbox{For } p=k+1, \; \alpha_{k+1}^{k+2}-\alpha_{k+1}^{k+1}-1=0	\nonumber \\
\Rightarrow \alpha_{k+1}^{k+1}(\alpha_{k+1}-1)=\alpha_{k}^{k}(\alpha_{k}-1) \nonumber \\
\Rightarrow \left(\frac{\alpha_{k}}{\alpha_{k+1}}\right)^{k}=\left(\frac{\alpha_{k+1}-1}{\alpha_{k}-1}\right).\alpha_{k+1}
\end{eqnarray}
\par From (\ref{eq_p}) we can argue,
\begin{itemize}
\item	$\alpha_{k}\neq\alpha_{k+1}$, since neither of them is $0$ or $1$ (from Lemma 2).
\item If $\alpha_{k}<\alpha_{k+1}$, we have LHS of inequality (\ref{eq_p}) $<1$, but RHS $>1$, since both 
the terms in RHS will be greater than 1 (by our assumption and by Lemma 2), a contradiction. 
\item Hence we must have 
	\begin{equation}
		\label{eq_p_1}
		\alpha_{k}>\alpha_{k+1}, \;\forall{k}\in\mathbb{N}
	\end{equation}
\end{itemize}
\par Also, from Lemma 2, we have, $1<\alpha_{k}<2,\;\forall{k}\in\mathbb{N}$.
\par Hence we have,
	\begin{eqnarray*}
		\label{eq_p_1_1}
		2>\alpha_{1}>\alpha_{2}>\ldots>\alpha_{k}>\alpha_{k+1}>\ldots>1, \;k\in\mathbb{N} \\
		\Rightarrow \lim_{k \rightarrow \infty}{\alpha_{k}}=1
	\end{eqnarray*}
	
\par Again, from (\ref{eq_p}) we have,
\begin{eqnarray}
\label{eq_p_2}
\Rightarrow \left(\frac{\alpha_{k+1}-1}{\alpha_{k}-1}\right).\alpha_{k+1}>1,
\; \mbox{since } \left(\frac{\alpha_{k}}{\alpha_{k+1}}\right)^k>1 \mbox{, from (\ref{eq_p_1})} \nonumber \\
\Rightarrow 2 > \alpha_{k+1} > \left(\frac{\alpha_{k}-1}{\alpha_{k+1}-1}\right) \mbox{, (from Lemma 2)} \nonumber \\
\Rightarrow \alpha_{k+1} > \frac{1+\alpha_{k}}{2}
\end{eqnarray}
\par Now, let us induct on $p \in \mathbb{N}$ to prove $\alpha_{p}^{p}<p+1$.
\par Base case: 
\par For $p=1$, $1<\alpha_{1}<2$, by Lemma 2.
\par Let us assume the inequality holds $\forall{p}\leq k \Rightarrow p<\alpha_{p}^{p}<p+1$, $\forall{p}\leq k$ 
\par Induction Step:
\begin{eqnarray}
	\label{eq_p_3}
	p=k+1, \;\; \alpha_{k+1}^{k+1}=\alpha_{k}^{k}.\left(\frac{\alpha_{k}-1}{\alpha_{k+1}-1}\right) 
	\mbox{, by (\ref{eq_p})} \nonumber \\
	\Rightarrow \alpha_{k+1}^{k+1}<(k+1).\left(\frac{\alpha_{k}-1}{\alpha_{k+1}-1}\right) \nonumber \\
	k<\alpha_{k}^{k}<k+1 \mbox{, by induction hypothesis} 					\nonumber \\
	\Rightarrow \alpha_{k+1}^{k+1}<(k+1).\left(1+\frac{\alpha_{k}-\alpha_{k+1}}{\alpha_{k+1}-1}\right) \nonumber \\
	\Rightarrow \alpha_{k+1}^{k+1}<(k+1)+\left(\frac{\alpha_{k}-\alpha_{k+1}}{\alpha_{k+1}-1}\right) \nonumber \\
	\Rightarrow \alpha_{k+1}^{k+1}<(k+1)+1, \; \left(\mbox{from (\ref{eq_p_2}), we have,}
	\frac{\alpha_{k}-\alpha_{k+1}}{\alpha_{k+1}-1}<1\right)	\nonumber \\
	\Rightarrow \alpha_{k+1}^{k+1}<(k+2) \nonumber \\
	\Rightarrow \alpha_{p}^{p}<(p+1) \mbox{, } \forall{p} \in \mathbb{N} 
\end{eqnarray}
Also, since $\alpha_{p}$ is a root of $x^{p+1}-x^{p}-1=0$, for $p=k$ we have.
\begin{eqnarray}
	\label{eq_p_4}
	\alpha_{k}^{k+1}-\alpha_{k}^{k}-1=0  \nonumber \\
	\Rightarrow \alpha_{k}^{k+1} = \alpha_{k}^{k} + 1 > 1 + 1\;\;(\mbox{since from Lemma 2, we have, }\alpha_{k}>1) \nonumber \\
	\Rightarrow \alpha_{k}^{k+1} > 2
\end{eqnarray}

\subsection{Lemma 4}
\label{sec:Lemma4}
\par The following inequalities always hold:
\begin{itemize}
\item $(k+1)^{\frac{1}{k}}<k^{\frac{1}{k-1}}<\ldots<4^{\frac{1}{3}}<3^{\frac{1}{2}}<2$
\item $\alpha_p^{p}<p+1\Rightarrow\alpha_p^{p-1}<p\Rightarrow\ldots\alpha_p^{3}<4\Rightarrow\alpha_p^{2}<3\Rightarrow\alpha_p<2$
\item $\alpha_p^{p+2}>2\Rightarrow\alpha_p^{p+3}>3\Rightarrow\ldots\alpha_p^{p+p}>p\Rightarrow\alpha_p^{p+p+1}>p+1$
\end{itemize}
\paragraph{Proof:}
\label{sec:Proof_4}
\par By Binomial Theorem, we have, 
\begin{eqnarray}
	\label{eq_bin}
(k+1)^{k-1}=\sum_{r=0}^{k-1}{\frac{\left(k-1\right)\left(k-2\right)\ldots\left(k-r\right)}{r!}.k^{k-1-r}} \nonumber \\
=\sum_{r=0}^{k-1}{\frac{\left(1-\frac{1}{k}\right)\left(1-\frac{2}{k}\right)\ldots\left(1-\frac{r}{k}\right)}{r!}.k^{k-1}}  
=\sum_{r=0}^{k-1}{\frac{\prod_{s=1}^{r}{\left(1-\frac{s}{k}\right)}}{r!}.k^{k-1}} \nonumber \\
<\underbrace{(1+1+1+..+1)}_{k\;times}.k^{k-1}=k.k^{k-1}=k^k 
\Rightarrow (k+1)^{\frac{1}{k}}<k^{\frac{1}{k-1}}
\end{eqnarray}  	
\par Hence we have, $(k+1)^{\frac{1}{k}}<k^{\frac{1}{k-1}}<\ldots<4^{\frac{1}{3}}<3^{\frac{1}{2}}<2$
\par Also, from (\ref{eq_p_3}) we have, $\alpha_{k}<(k+1)^{\frac{1}{k}}$. 
\par Combining, we get, 
\begin{eqnarray}
\alpha_{k}<(k+1)^{\frac{1}{k}}<k^{\frac{1}{k-1}}<\ldots<4^{\frac{1}{3}}<3^{\frac{1}{2}}<2 \nonumber \\
\alpha_{k}^{k}<(k+1)\Rightarrow\alpha_{k}^{k-1}<k\ldots \Rightarrow \alpha_{k}^{4}<5 \Rightarrow \alpha_{k}^{3}<4 \Rightarrow \alpha_{k}^{2}<3 \Rightarrow \alpha_{k}<2  
\end{eqnarray}

Also, we have,
\begin{eqnarray}
\alpha_{k}^{k+1}>2\Rightarrow\alpha_{k}^{k+2}=\alpha_{k}^{k+1}+\alpha_{k}>2+1=3 \nonumber \\
\alpha_{k}^{k+2}>3\Rightarrow\alpha_{k}^{k+3}=\alpha_{k}^{k+2}+\alpha_{k}^{2}>3+1=4 \nonumber \\
\Rightarrow \alpha_p^{p+2}>2\Rightarrow\alpha_p^{p+3}>3\Rightarrow\ldots\alpha_p^{p+p}>p\Rightarrow\alpha_p^{p+p+1}>p+1
\end{eqnarray}

\subsection{Lemma 5}
\label{sec:Lemma5}
\par The following inequality gives us the lower and upper bounds for generalized Fibonacci-p-sequence,
\begin{equation}
\alpha_{p}^{n}>F_{p}(n)>\alpha_{p}^{n-p},\;\forall{n}>p,\;n \in \mathbb{N}
\label{fibpBound}
\end{equation}
\par where $\alpha_{p}$ is the $positive$ root of the equation $x^{p+1}-x^{p}-1=0$.
\paragraph{Proof:} 
\label{sec:Proof_5}
\par We induct on $n$ to show the result. \\
\par $F_{p}(0)=F_{p}(1)=\ldots=F_{p}(p)=1$, (By definition of Fibonacci-p-Sequence). \\
\par Base case: 
\par From Lemma 4, we have, 
\begin{eqnarray*}
\alpha_{p}^{p+1}>F_{p}(p+1)=F_{p}(p)+F_{p}(0)=1+1=2>\alpha_{p} \\
\alpha_{p}^{p+2}>F_{p}(p+2)=F_{p}(p+1)+F_{p}(1)=2+1=3>\alpha_{p}^{2} \\
\alpha_{p}^{p+3}>F_{p}(p+3)=F_{p}(p+2)+F_{p}(2)=3+1=4>\alpha_{p}^{3} \\
\cdots \;\;\;\;\;\;\;\;\; \cdots \;\;\;\;\;\;\;\;\; \cdots \;\;\;\;\;\;\;\;\; \\
\alpha_{p}^{p+p+1}>F_{p}(p+p+1)=F_{p}(p+p)+F_{p}(p)=(p+1)+1=p+2>\alpha_{p}^{p+1} \\
\label{fibpprf_1}
\end{eqnarray*}
\par Induction Step:
\par Let's assume the above result is also true $\forall{m}: (2p+1<m<n),\;m,n\in \mathbb{N}$. Now, we prove for $m=n$,
\begin{eqnarray*}
\alpha_{p}^{n-1}+\alpha_{p}^{n-p-1}>F_{p}(n-1)+F_{p}(n-p-1) \; (by \;\;hypothesis)\\
F_{p}(n-1)+F_{p}(n-p-1)>\alpha_{p}^{n-p-1}+\alpha_{p}^{n-2p-1} \; (by \;\;hypothesis)\\
\Rightarrow \alpha_{p}^{n-p-1}.(1+\alpha_{p}^{p})>F_{p}(n)>\alpha_{p}^{n-2p-1}.(1+\alpha_{p}^{p}) \\
\Rightarrow \alpha_{p}^{n-p-1}.\alpha_{p}^{p+1}>F_{p}(n)>\alpha_{p}^{n-2p-1}.\alpha_{p}^{p+1} \\
\Rightarrow \alpha_{p}^{n}>F_{p}(n)>\alpha_{p}^{n-p},\;\forall{n}>p,\;n \in \mathbb{N}
\label{fibpprf}
\end{eqnarray*}

Hence we have the following inequality,

\begin{eqnarray}
\label{lb:}
(\alpha_{p})^{n}>F_{p}(n)>(\alpha_{p})^{n-p}, \; \alpha_{p}\in\;\Re^{+}\; and\; \alpha_{p}\in\;(1,2) \\
\nonumber \alpha_{1}=\frac{1+\sqrt{5}}{2}\approx 1.618034, \\ 
\nonumber \alpha_{2} \approx 1.465575, \\
\nonumber \alpha_{3} \approx 1.380278, \\
\nonumber \alpha_{4} \approx 1.324718, \\
\nonumber \alpha_{p}>\alpha_{p+1},\;\forall{p}\in \mathbb{N} 
\end{eqnarray}
\par The empirical results (Table 1) also prove our claim for $p=2$. \\

\par
\begin{figure}[htbp]
	\label{fig:t1}
	\centering
		\includegraphics[width=12cm,height=7.5cm]{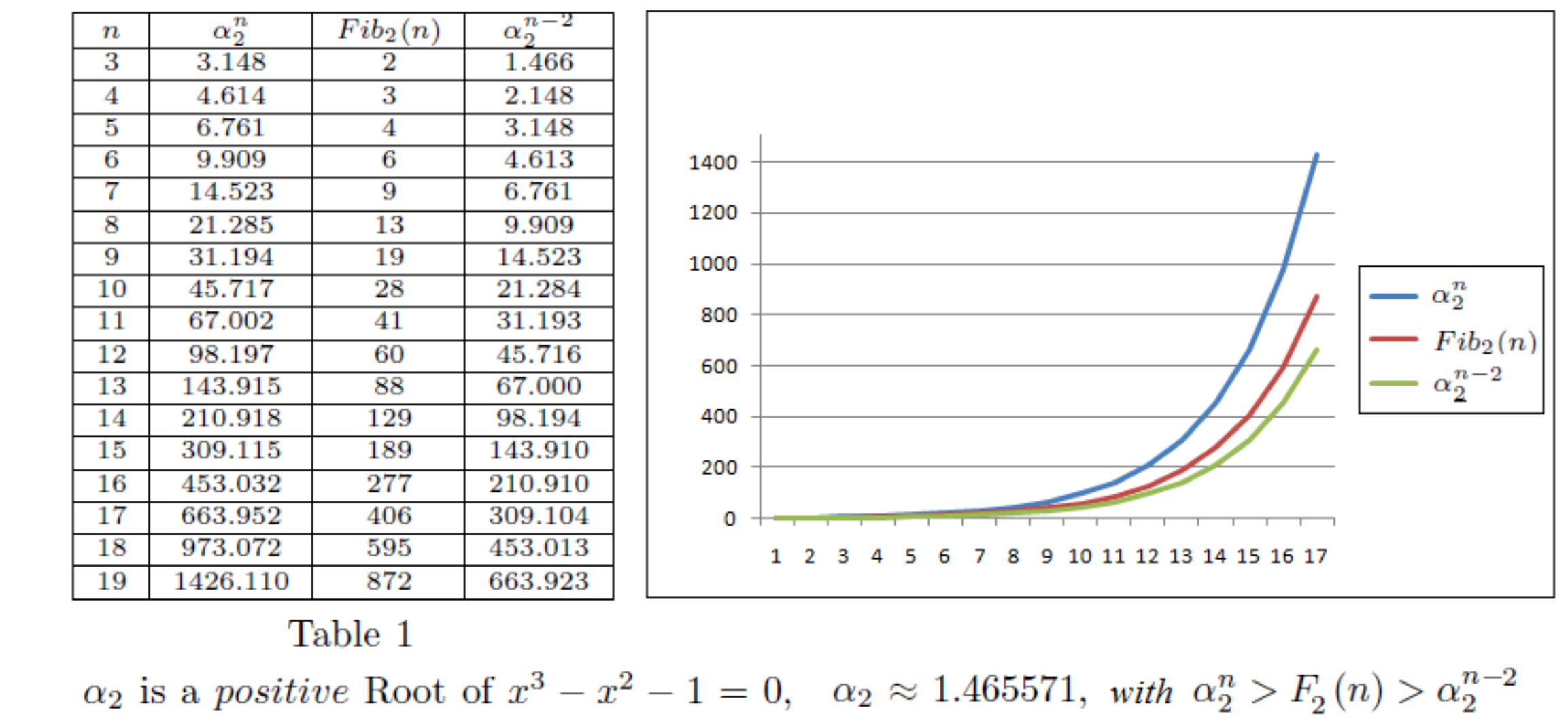}
\end{figure}

\par Also $F_{p}(0)=F_{p}(1)=\cdots=F_{p}(p)=1$ and $F_{p}(n+1)>F_{p}(n), \forall{n}>p$. Hence we have,

\begin{eqnarray}
\label{ub:}
\nonumber F_{p}(n)=F_{p}(n-1)+F_{p}(n-p-1)<2.F_{p}(n-1),\; \forall{n}>p  \\
\nonumber \Rightarrow F_{p}(n)<2.F_{p}(n-1)<2^{2}.F_{p}(n-2)<\cdots<2^{n-p}.F_{p}(p)=2^{n-p},\; \forall{n}>p  \\
\Rightarrow F_{p}(n) < 2^{n-p}, \; \forall{n}>p \;\;
\end{eqnarray}

\par Combining (\ref{lb:}) and (\ref{ub:}), we have,
\begin{eqnarray}
\label{bounds:}
(2)^{n-p} > F_{p}(n) > (\alpha_{p})^{n-p}, \; \forall{n}>p, \; and\; n,p \in \mathbb{N}
\end{eqnarray}
\par where $\alpha_{p}$ is the positive Root of $x^{p+1}-x^{p}-1=0$. \\

Figure 2 and Table 2 show the convergence of ratio of successive terms for Fibonacci-p-sequences for different $p$ (For $p=1$ we get classical Fibonacci sequence). It can be noticed that smaller the value of $p$, quicker the convergence of the ratio is achieved, as shown. Also value to which the ratio converges monotonically decreases with increase in the value of $p$.

\begin{figure}[htbp]
	\label{fig:f2}
	\centering
		\includegraphics[width=12cm,height=15cm]{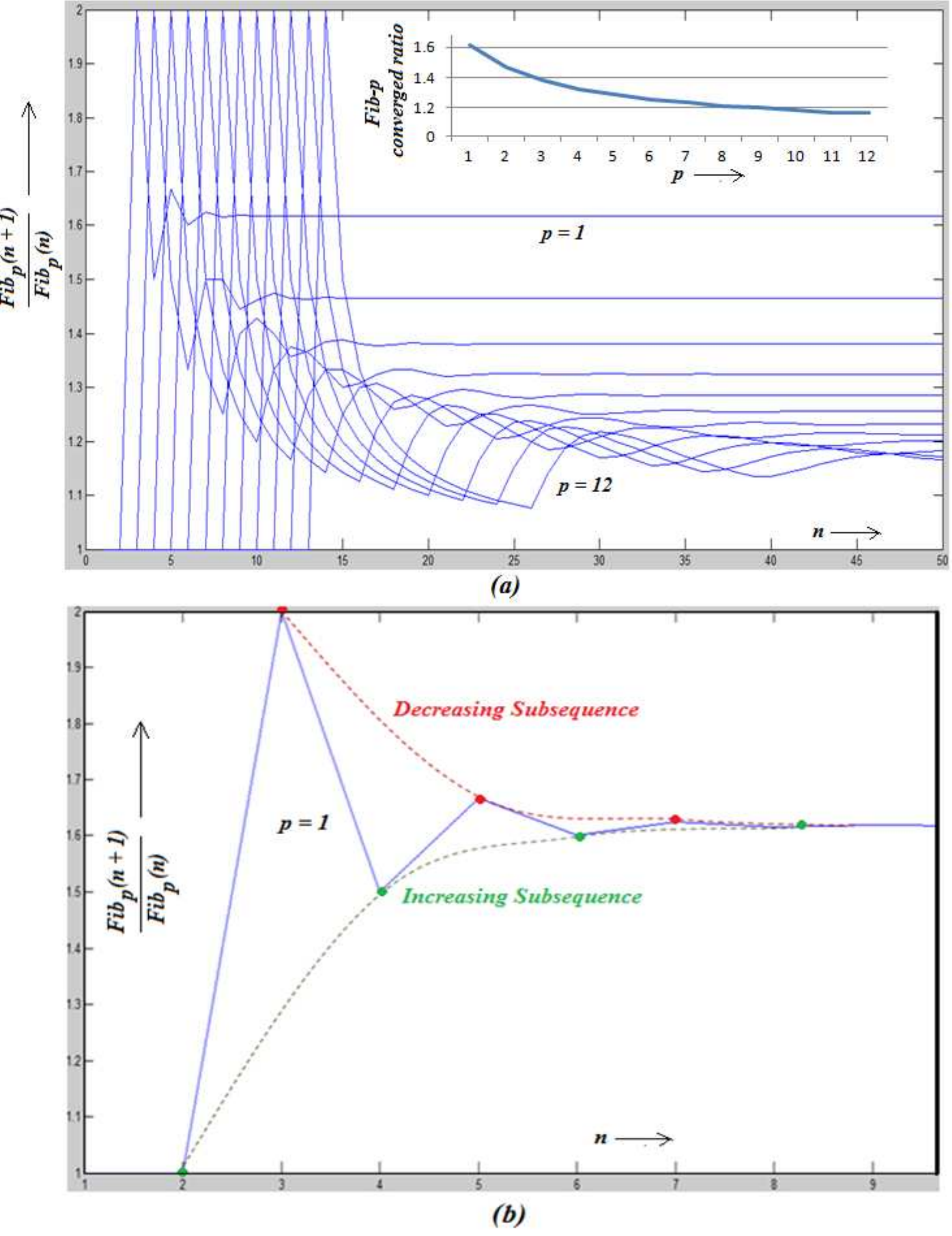}
	\caption{(a) Convergence of the ratio of successive terms in generalized Fibonacci p-Sequence for different values of $p$
					 (b) Convergent Subsequences for classical Fibonacci sequence (p=1)}
\end{figure}

\begin{figure}[htbp]
	\label{fig:t2}
	\centering
		\includegraphics[width=7.5cm,height=12.5cm]{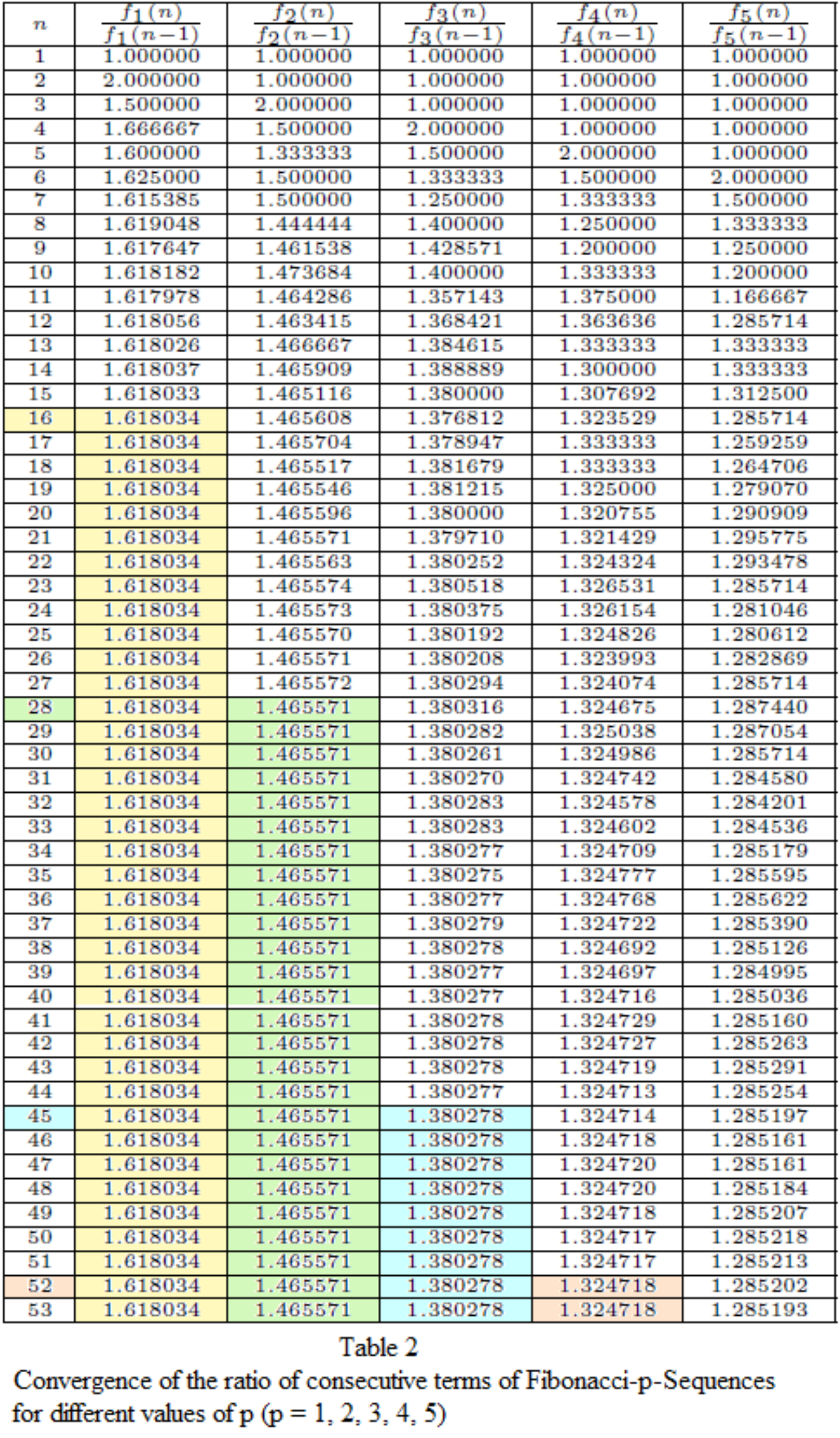}
\end{figure}

\section{Application in Data Hiding}
\par Data hiding is a new kind of secret communication technology, where message is hidden inside 
an image or any other medium, so that it cannot be observed. One of the simplest data hiding technique
is LSB data hiding technique, which merely embeds secret message-bits in a subset of the LSB planes of 
the image. One of the drawbacks of this technique is: as soon as we go from LSB to MSB for selection of 
bit-planes for our message embedding, the distortion in stego-image is likely to increase
exponentially, so it becomes impossible (without noticeable distortion and with exponentially increasing 
distance from cover-image and stego-image) to use higher bit-planes for embedding without any further 
processing.

\par This particular problem was addressed by Battisti {\it et al.}, \cite{battisti06}, who proposed to use generalized 
Fibonacci-p-sequence decomposition technique instead of classical binary decomposition and shows by empirical 
results that this technique outperforms the classical LSB technique when thought in terms of embedding in the 
higher bit plane as well with less distortion. This technique basically increases the number of bit-planes (by generating
a new larger set of bit-planes that we call virtual bit-planes) by using Fibonacci-p-sequence decomposition. It can 
be further improved using prime and natural number decomposition techniques as shown in \cite{dey07a}, \cite{dey07b}, 
\cite{dey08}. Also, \cite{cooper84} \cite{dotson93} illustrates Fibonacci sequence can be used in various applications.

\par In this paper we give a formal proof of Fibonacci-p-sequence bounds and show how this can be used to theoretically 
prove that Fibonacci-p-sequence decomposition gives better result in hiding data. From (\ref{bounds:}) it is clear 
(from the upper bound) that the same value will require more numbers of bits to be represented than the number of bits required in classical binary decomposition (since $2^n>\alpha^n>F_p{n}$), if it's expressed using Fibonacci-p-sequence decomposition (where the radix is Fibonacci-p-sequence numbers instead of powers of $2$). \\

As illustrated in \cite{dey08}, in order to measure the distortion in the stego-image, we use Mean square error (MSE),
Worst case Mean Square Error (WMSE) and Peak Signal to Noise Ratio (PSNR), which are defined by
\begin{eqnarray*}
MSE = \displaystyle\sum_{i=1}^{M}\sum_{j=1}^{N}(f_{ij}-g_{ij})^2/MN \\
PSNR = 10.log_{10}\left(\frac{L^2}{MSE}\right) 	
\end{eqnarray*} \cite{dey08}.
If the secret data-bit is embedded in the $i^{th}$ bit-plane of a pixel, the worst-case
error-square-per-pixel will be $= WSE = |W(i)(1..0)|^2 = (W(i))^2$ (here $W(i)$ represents the corresponding weight in the number system for the $i^{th}$ bit, e.g., for classical decomposition $W(i)=2^i$, for generalized Fibonacci decomposition $W(i)=F_p{i}$), corresponding to the case when the corresponding bit in cover-image toggles in stego-image, after embedding the secret data-bit. For example, worst-case error-square-per-pixel for embedding a secret data-bit in the ith bit plane in case of a pixel in classical
binary decomposition is $= (2^i)^2 = 4^i$, where $i \in \mathbb{N} \cup \{0\}$. If the original $k$-bit
grayscale cover-image has size $w \times h$, we define, $WMSE = w \times h \times (W(i))^2 = w \times h \times WSE$ \cite{dey08}. 
Hence,
\par WMSE after embedding secret message bit only in the $l^{th}$ (virtual) bit-plane of each pixel in case of classical 
(traditional) binary (LSB) data hiding technique is given by,
\begin{eqnarray*}
    {\left({WMSE}_{l^{th} bit-plane}\right)}_{Classical-Binary-Decomposition}=\theta(4^{l}). 
    \label{wmselsb}
\end{eqnarray*}

\par WMSE after embedding secret message bit in the $l^{th}$ (virtual) bit-plane of each pixel in case of 
generalized Fibonacci decomposition is given by,

 \begin{eqnarray*}
 {\left({WMSE}_{l^{th}\;bit-plane}\right)}_{Fibonacci-p-Sequence\;Decomposition}
 =\left(F_p(l)\right)^{2} \\
 \Rightarrow \left(\alpha_{p}^{2}\right)^{l-p} < {\left({WMSE}_{l}\right)}_{Fibonacci-p-sequence} 
 <\left(\alpha_{p}^{2}\right)^{l}, \\
 \alpha_{p}\in\Re^{+},\;\alpha_{1}=\frac{1+\sqrt{5}}{2},\\
 \alpha_{p}^{2}>\alpha_{p+1}^{2},\forall{p}\in \mathbb{N}, \alpha_{1}^{2} \approx 2.618, \\
 \Rightarrow {\left({WMSE}_{l}\right)}_{Generalized-Fibonacci-p-sequence}  < \theta \left({2.618}^l \right).
 \end{eqnarray*}

\par Hence, we have,
\begin{eqnarray*}
{\left(WMSE\right)}_{Binary} >{\left(WMSE\right)}_{Fibonacci} \\
\Rightarrow {\left(PSNR\right)}_{Fibonacci} > {\left(PSNR\right)}_{Binary}.
\end{eqnarray*}

\par Thus, we have first proved bounds on the generalized Fibonacci sequence and then by using our bounds, 
we have given a formal proof for better performance (in terms of PSNR) for LSB data hiding technique using 
generalized Fibonacci-p-sequence decomposition than that using classical binary decomposition. Also, we have,
\begin{eqnarray*}
(\mbox{number of primes} \leq n) &=& p_{n}=\theta{(n.\log{n})} \\
&=& o(\alpha_{p})^{n-p} < F_{p}(n) < 2^{n-p}
\end{eqnarray*}
{ (\cite{telang99}, \cite{niven66}, \cite{tatt05})}
\par The above implies that if the same number is represented using prime decomposition ($n^{th}$ prime number as weightage 
to $n^{th}$ bit), it will give still more numbers of virtual bit planes \cite{battisti06}. 

\par We have similar results for LSB data hiding using natural number decomposition technique, and combining the results 
(from \cite{battisti06}, \cite{dey07a}, \cite{dey07b}, \cite{dey08}) we have the following,
\begin{eqnarray*}
{\left(WMSE\right)}_{Binary} >{\left(WMSE\right)}_{Fibonacci} > {\left(WMSE\right)}_{Prime}>{\left(WMSE\right)}_{Natural}\\
\Rightarrow {\left(PSNR\right)}_{Natural}>{\left(PSNR\right)}_{Prime}>{\left(PSNR\right)}_{Fibonacci} > {\left(PSNR\right)}_{Binary}.
\end{eqnarray*}

\par Hence, data hiding using natural number decomposition gives the best performance among the above mentioned techniques.

\par In data hiding, we hide data in different bit-planes of a pixel. In classical LSB data-hiding technique the 
pixel is represented as binary value, hence it has less numbers of bit-planes as we have in case of Fibonacci-p-sequence 
decomposition, the later having still less number of bit-planes than in case of prime decomposition technique.
It is shown in \cite{dey07a}, \cite{dey07b}, \cite{dey08} by calculation of WMSE and PSNR measures that embedding data even 
in higher bit-planes of pixel using these techniques results in less visible distortion of the cover image, since the distortion
as measured by WMSE is proportional to the square of the weights to the bits in the corresponding decomposition, hence it decreases
as the weights go on decreasing from classical binary to generalized Fibonacci and from that to prime and natural number decomposition \cite{dey08}.

\section{Conclusions}
In this paper, we have established the bounds for generalized Fibonacci-sequence $(\alpha)^{n} > F_{p}(n) > (\alpha_{p})^{n-p}, \; \forall{n}>p, (n,p) \in \mathbb{N}$. Empirical results obtained vindicates our theoretically-proven bounds. Then we used the 
result $(2)^{n-p} > F_{p}(n) > (\alpha_{p})^{n-p}, \; \forall{n}>p, (n,p) \in \mathbb{N} $, where 
$\alpha_{p}$ is the positive Root of $x^{p+1}-x^{p}-1=0$, to prove that data hiding technique using generalized 
Fibonacci-p-sequence gives more embeddable bit-planes along with better PSNR than that in case of 
classical LSB technique \cite{dey08}, and the same using prime decomposition technique increases virtual bit-planes 
and PSNR further \cite{dey07a}.

\end{document}